\RequirePackage{silence}
\documentclass[twocolumn,10pt,groupedaddress,nofootinbib,aps,prl]{revtex4-2} 
\pdfoutput=1

\usepackage[utf8]{inputenc}
\usepackage{hyperref}
\usepackage{braket}
\usepackage{upgreek}
\usepackage[normalem]{ulem}
\usepackage[dvips]{graphicx}
\usepackage{hhline}
\usepackage{epsfig,amsmath,amssymb,verbatim,mathrsfs,array,layout,textcomp,latexsym,slashed,graphicx,booktabs,color,mathtools}
\usepackage[caption=false]{subfig}

\usepackage{microtype} 
\begin{document}

\title{Pairwise Helicity in Higher Dimensions}

\author{Yale Fan}
\affiliation{Weinberg Institute for Theoretical Physics, Department of Physics, University of Texas at Austin, Austin, TX 78712, USA} 

\begin{abstract}
Studies of scattering amplitudes for electric and magnetic charges have identified previously overlooked multiparticle representations of the Poincar\'e group in four dimensions. Such representations associate nontrivial quantum numbers (known as pairwise helicities) with asymptotically separated pairs of particles, and thus cannot be described as tensor products of one-particle states. We extend this construction to sources and spacetimes of higher dimension. We first establish the dynamical origin of pairwise helicity in $p$-form electrodynamics coupled to mutually nonlocal branes. We then interpret this pairwise helicity as a quantum number under an $SO(2)$ pairwise little group associated with pairs of distinct branes. We further characterize the ``higher'' little groups that could in principle be used to induce multiparticle or multi-brane representations of the Lorentz group.
\end{abstract}

\maketitle

\section{Introduction}

A foundational assumption in most discussions of the $S$-matrix in relativistic quantum field theory is that asymptotic states should factorize as tensor products of single-particle states, as classified by Wigner \cite{Wigner:1939cj}.  However, this assumption breaks down for the scattering of particles with both electric and magnetic charge, for which the angular momentum in the electromagnetic field remains finite even at asymptotic separation \cite{Zwanziger:1972sx}.  This effect is codified by the notion of pairwise helicity, or the charge under the pairwise little group, defined as the subgroup of the Lorentz group that leaves the momenta of a given pair of particles invariant \cite{Csaki:2020inw, Csaki:2020yei}.  This group has nontrivial representations on certain multiparticle states, leading to fundamentally new multiparticle representations of the Poincar\'e group in four dimensions.

The pairwise little group naturally generalizes to the ``$n$-particle'' little group for $n > 2$, which becomes nontrivial in sufficiently high dimension.  One might wonder whether such a group can be represented nontrivially in any physical theory.  While this compelling question remains open, we show in this paper that there does exist a physically manifest generalization of the pairwise little group to higher dimensions: the pairwise little group of mutually nonlocal branes.

To motivate this construction, we first recall how the pairwise helicity of four-dimensional multiparticle states arises dynamically from the electromagnetic angular momentum of dyons.  While irreducible single-particle representations of the Poincar\'e group are induced by representations of the corresponding little group, representations of the pairwise little group induce multiparticle representations that go beyond tensor products of single-particle representations.

We next compute the angular momentum in the electromagnetic field between mutually nonlocal branes of arbitrary (complementary) dimensions.  It admits a simple expression in terms of the Dirac-Schwinger-Zwanziger (DSZ) pairing \cite{Dirac:1931kp, Schwinger:1969ib, Zwanziger:1968rs} within $p$-form electrodynamics.  To provide a framework for this higher-dimensional analogue of pairwise helicity, we define representations of the Lorentz group that describe infinitely extended, isotropic branes. (For reasons that will become clear, such representations do not generally lift to representations of the Poincar\'e group.) A $(p - 1)$-brane representation of the Lorentz group in $d$ spacetime dimensions is induced by a representation of the corresponding little group, which contains the group of transverse rotations $SO(d - p)$.  Multi-brane representations can further carry nontrivial quantum numbers under little groups associated with pairs of distinct branes.  Such pairwise little groups include rotations in the directions mutually transverse to both branes.  When this subgroup of mutually transverse rotations is $SO(2)$, we refer to the associated charge as the pairwise helicity.

Finally, we offer some thoughts on the higher ``$n$-brane'' little groups that could in principle induce more general multi-brane representations of the Lorentz group.

\section{Pairwise Helicity for Particles}

Our analysis of $(p - 1)$-branes in $d$ dimensions builds on the familiar case $d = 4$, $p = 1$.  Recall that particles (0-branes) furnish unitary irreducible representations of the Poincar\'e group.  Such representations can be classified by their mass and their representation with respect to the little group, the subgroup of the Lorentz group that leaves some reference momentum $k$ of that mass invariant \cite{Wigner:1939cj}.  A general Lorentz transformation $\Lambda$ acts on a single-particle momentum eigenstate $|p; \sigma\rangle$ as
\begin{equation}
U(\Lambda)|p; \sigma\rangle = D_{\sigma'\sigma}(W)|\Lambda p; \sigma'\rangle,
\end{equation}
where $W = W(\Lambda, p, k)$ is a little group transformation, $D$ is a unitary representation of $W$, and the single-particle quantum number $\sigma$ labels a little group representation.  A particle representation of the Poincar\'e group (and in particular, of the Lorentz group) is thus \emph{induced} by a unitary irreducible representation of the little group.

On the other hand, multiparticle representations of the Poincar\'e group may carry additional quantum numbers that characterize the Lorentz transformation properties of multiparticle states relative to tensor products of one-particle states \cite{Csaki:2020inw, Csaki:2020yei}.  Given a generic pair of reference four-momenta (either massive or massless), the pairwise little group is defined as the $SO(2)$ subgroup of the Lorentz group that preserves both.  Since three-particle and higher little groups are trivial in four dimensions, a general $n$-particle state takes the form
\begin{equation}
|p_1, \ldots, p_n; \sigma_1, \ldots, \sigma_n; q_{12}, \ldots, q_{n-1, n}\rangle,
\end{equation}
where $p_i$ is the momentum of particle $i$, the $\sigma_i$ are single-particle quantum numbers (spins or helicities), and the $\binom{n}{2}$ quantized pairwise helicities $q_{ij}$ are associated with pairs of particles $i, j$.  Under a Lorentz transformation $\Lambda$, these states may transform with extra phases arising from nontrivial pairwise little group representations:
\begin{align}
&U(\Lambda)|p_1, \ldots, p_n; \sigma_1, \ldots, \sigma_n; q_{12}, \ldots, q_{n-1, n}\rangle = \prod_{i<j} e^{iq_{ij}\phi_{ij}} \nonumber \\
&\times \prod_{i=1}^n D_{\sigma_i'\sigma_i}^i|\Lambda p_1, \ldots, \Lambda p_n; \sigma_1', \ldots, \sigma_n'; q_{12}, \ldots, q_{n-1, n}\rangle.
\end{align}
The angles $\phi_{ij} = \phi(\Lambda, p_i, p_j)$ parametrize pairwise little group transformations, while the $D$'s are unitary representations of one-particle little group transformations.

There is precisely one setting in which nontrivial pairwise helicities are known to arise: the dynamics of electric and magnetic charges.  If particles $i, j$ are dyons of charges $(e_i, g_i)$ and $(e_j, g_j)$, then $q_{ij}$ is simply the quantized angular momentum $\frac{1}{4\pi}(e_i g_j - e_j g_i)$ in their mutual electromagnetic field.  This is easiest to see in a Lorentz frame where their spatial momenta are equal and opposite.  Indeed, for a spinless two-dyon state, $q_{ij}$ is the charge under rotations about the axis of the two asymptotically separated dyons.  A careful derivation of this result for non-static sources is given in \cite{Zwanziger:1972sx}.

\section{Pairwise Helicity for Branes}

The above discussion makes clear that pairwise helicity in four dimensions is a reflection of the mutual nonlocality of the underlying electromagnetic sources.

A natural starting point for generalizing pairwise helicity to higher dimensions is to consider $p$-form electrodynamics in $d$-dimensional Minkowski space.  In this theory, an electrically charged $(p - 1)$-brane couples directly to the $p$-form gauge potential $A$.  On the other hand, a magnetic $(d - p - 3)$-brane \cite{Nepomechie:1984wu, Teitelboim:1985yc, Henneaux:1986ht} can be thought of as coupling to $A$ through a $(d - p - 1)$-form current $G$ localized to the worldvolume of a $(d - p - 2)$-dimensional Dirac brane (or generalized Dirac string \cite{Dirac:1931kp}) that it bounds.  The total $(p + 1)$-form field strength is then given by $F = dA + \ast G$.

Let us label the time coordinate of $\mathbb{R}^{d - 1, 1}$ by $x_0$ and the spatial coordinates by
\begin{equation}
(\vec{x}, \vec{y}, \vec{z})\equiv (x_1, x_2, x_3, y_1, \ldots, y_{p-1}, z_1, \ldots, z_{d - p - 3}).
\label{coordinatelabels}
\end{equation}
Suppose we have an electric $(p - 1)$-brane lying along $\vec{y}$ and a magnetic $(d - p - 3)$-brane lying along $\vec{z}$.  For a magnetic brane (``$i$'') of charge density $g_i$ and an electric brane (``$j$'') of charge density $e_j$ in a configuration such as that of \eqref{coordinatelabels}, we define their pairwise helicity to be
\begin{equation}
q_{ij} = \frac{1}{4\pi}\begin{cases} (-1)^p e_j g_i & \text{if $d\neq 2(p + 1)$}, \\ e_i g_j + (-1)^p e_j g_i & \text{if $d = 2(p + 1)$}. \end{cases}
\label{defpairwise}
\end{equation}
Note that the branes can be dyonic when $d = 2(p + 1)$, in which case we additionally assign the magnetic brane an electric charge density $e_i$ and the electric brane a magnetic charge density $g_j$.

The quantity $q_{ij}$ in \eqref{defpairwise} is quantized as a half-integer.  This statement is precisely the Dirac quantization condition when $d\neq 2(p + 1)$ \cite{Nepomechie:1984wu, Teitelboim:1985yc, Henneaux:1986ht, Wu:1975es} and the more refined DSZ quantization condition when $d = 2(p + 1)$ \cite{Bremer:1997qb, Deser:1997se, Bertolini:1998mg, Deser:1998vc, Bekaert:2002cz}.  Indeed, the trajectories of electric and magnetic branes can link in $(d - 1)$-dimensional space.  Hence amplitudes involving these objects incur an Aharonov-Bohm phase, and quantum-mechanical consistency of the theory places constraints on the allowed electric and magnetic charges.

More directly, however, $q_{ij}$ is nothing other than the angular momentum in the electromagnetic field of branes $i$ and $j$.  To see this, we focus on the spatial directions $\mathbb{R}^{d-1}$ in \eqref{coordinatelabels}.  We take brane $i$ (which fills the $\vec{z}$-directions) to lie at the origin in $\vec{x}$, while we take brane $j$ (which fills the $\vec{y}$-directions) to lie at the point $\smash{\vec{R}} = (0, 0, R_3)$ in $\vec{x}$.  The angular momentum corresponding to rotations in the $x_1, x_2$ directions is
\begin{equation}
J_{12} = \int d^{d-1} x\, (x_1 T^0{}_2 - x_2 T^0{}_1),
\end{equation}
where $T^0{}_i = \frac{1}{p!}F^0{}_{i_1\cdots i_p}F_{i}{}^{i_1\cdots i_p}$ is the linear momentum density of the gauge field.  We can write the electric and magnetic fields sourced by the branes as differential forms $E$ and $B$ on $\mathbb{R}^{d-1}$, with components $E^{i_1\cdots i_p} = F^{0i_1\cdots i_p}$ and $B^{i_1\cdots i_{d-p-2}} = (-1)^p(\ast F)^{0i_1\cdots i_{d-p-2}}$.  They satisfy
\begin{equation}
d\ast E = (-1)^{p-1}e_j\widehat{\Sigma}_j, \quad d\ast B = (-1)^{d-p-3}g_i\widehat{\Sigma}_i,
\label{intermediate}
\end{equation}
where $\smash{\widehat{\Sigma}_{i, j}}$ denote the Poincar\'e duals of the brane submanifolds $\Sigma_{i, j}\subset \mathbb{R}^{d-1}$ (the operations $d$, $\ast$, and $\widehat{\phantom{m}}$ in \eqref{intermediate} are all defined with respect to $\mathbb{R}^{d-1}$).  Using \eqref{intermediate}, we find:
\begin{equation}
J_{12} = \frac{(-1)^p e_j g_i}{4\pi}\frac{R_3}{|R_3|},
\label{theresult}
\end{equation}
which exactly reproduces \eqref{defpairwise}.  When $d = 2(p + 1)$, the $e_i g_j$ term in \eqref{defpairwise} follows from a symmetry argument.  Crucially, the result \eqref{theresult} is independent of the separation $|\smash{\vec{R}}| = |R_3|$, so it holds equally well for branes that are asymptotically separated in the $x_3$-direction.\footnote{Our mathematical conventions, details on the derivation of \eqref{theresult}, and additional context on $p$-form electrodynamics can be found in Appendices \ref{app:conventions}, \ref{app:angularmomentum}, and \ref{app:pform}.}

Having explained the dynamical meaning of $q_{12}$, we would like to show that it deserves the moniker ``pairwise helicity,'' i.e., to interpret it as an $SO(2)$ pairwise little group charge.  We propose that states of multiple branes can transform with nontrivial pairwise helicities when they contain pairs of mutually nonlocal branes with $SO(2)$ pairwise little groups.  To formalize this statement, we now suggest a framework in which to understand pairwise helicity more systematically.

\section{Brane Representations of the Lorentz Group}

It is interesting to ask whether branes, like particles, can be described as representations of an appropriate group of spacetime symmetries.  Since branes have infinitely many more degrees of freedom than particles, making such a question tractable requires certain simplifying assumptions.  For this reason, we concentrate on $(p - 1)$-branes of positive tension in $d$ spacetime dimensions that are completely isotropic \cite{Carter:1992ny}.  The tangent space to a point on the brane worldvolume is a $p$-dimensional linear subspace of Minkowski space that contains a timelike direction.  This ``brane subspace'' is characterized by a decomposition
\begin{equation}
\eta^{\mu\nu} = \Pi^{\mu\nu} + (\Pi^\perp)^{\mu\nu},
\end{equation}
where $\Pi$ and $\Pi^\perp$ are orthogonal projection operators onto the parallel and transverse directions.  They have rank $p$ and $d - p$, respectively.  The projection tensor $\Pi$ is manifestly invariant under internal $SO(p - 1, 1)$ Lorentz transformations, i.e., independent of the choice of basis for the brane subspace.

In a given $SO(d - 1, 1)$ Lorentz frame, the relativistic $d$-velocity of the brane is the normalized timelike vector in this subspace given by
\begin{equation}
u^\mu\equiv \frac{\Pi^{0\mu}}{\sqrt{\Pi^{00}}}.
\label{dvelocity}
\end{equation}
We use ``mostly minus'' signature, so that $u_\mu u^\mu = 1$.  The spatial orientation of the brane is then characterized by a spatial projection tensor $\Xi$ of rank $p - 1$:
\begin{equation}
\Xi^{\mu\nu}\equiv \Pi^{\mu\nu} - u^\mu u^\nu.
\label{spatialprojection}
\end{equation}
In any Lorentz frame, the brane's $d$-velocity is parallel to its worldvolume directions ($\Pi^\mu{}_\nu u^\nu = u^\mu$) but orthogonal to its $p - 1$ purely spatial directions ($\Xi^\mu{}_\nu u^\nu = 0$).

Note that, despite our notation, $u$ and $\Xi$ do not generally transform as Lorentz tensors.  On the other hand, because the brane orientation $\Pi$ transforms as a symmetric two-index Lorentz tensor, we can use it to construct representations of the Lorentz group $SO(d - 1, 1)$.  First, we observe that the subgroup of the Lorentz group that leaves $\Pi$ invariant is $SO(p - 1, 1)_\parallel\times SO(d - p)_\perp$, corresponding to rotations or boosts in the directions parallel ($\parallel$) and transverse ($\perp$) to the brane.  Indeed, we can always choose a Lorentz frame in which the worldvolume projection tensor takes the standard form
\begin{equation}
\tilde{\Pi}^{\mu\nu} = \operatorname{diag}(\underbrace{1, -1, \ldots, -1}_p, \underbrace{0, \ldots, 0}_{d - p}).
\label{referencepi}
\end{equation}
This can be achieved by boosting to the brane's rest frame, where $\tilde{u}^\mu = (1, 0, \ldots, 0)$, and performing an appropriate spatial rotation.  For simplicity, we consider only brane representations of the Lorentz group in which the noncompact $SO(p - 1, 1)_\parallel$ factor is represented trivially.  We refer to the compact $SO(d - p)_\perp$ factor as the single-brane little group, dropping the subscript $\perp$ from here on.  To specify a representation of the $SO(d - p)$ little group, we introduce a collective label $\sigma$ that includes all necessary Casimirs.  We call $\sigma$ the ``spin'' under $SO(d - p)$.

Following Wigner, we introduce a collection of single-brane states $|\Pi; \sigma\rangle$ that transform in an infinite-di\-men\-sion\-al unitary representation of the $d$-dimensional Lorentz group according to
\begin{equation}
U(\Lambda)|\Pi; \sigma\rangle = D_{\sigma'\sigma}(W)|\Lambda\Pi\Lambda^{-1}; \sigma'\rangle,
\end{equation}
where $W = W(\Lambda, \Pi, \tilde{\Pi})$ is a little group transformation. (Here, we write $\Pi$ for the matrix $\Pi^\mu{}_\nu$ without indices.) These brane states, unlike particle states, are not constructed as eigenstates of the translation generators of the Poincar\'e algebra.  Indeed, while the $d$-momentum per unit volume is given by $p^\mu = Tu^\mu$ in terms of the brane tension $T$ (with mass dimension $[T] = p$), $p^\mu$ does not correspond to the integrated Poincar\'e charge $P^\mu$ (which diverges in states of infinite branes), nor does it transform as a Lorentz vector.\footnote{Further details on this construction and its subsequent generalization to multiple branes are given in Appendix \ref{app:pairwise}.}

In the case of a particle ($p = 1$), we have $T = m > 0$ and $\Pi^{\mu\nu} = u^\mu u^\nu$.  Hence $\Pi$ and $u$ contain precisely the same information, under the assumption that the timelike component of $u$ is positive.  In this case, \eqref{dvelocity} becomes a tautology and $u$ transforms as a genuine Lorentz vector.  The little group reduces to the expected $SO(d - 1)$ for a massive particle.  In the opposite case of a space-filling brane ($p = d$), we have $\Pi^{\mu\nu} = \eta^{\mu\nu}$ in any frame, and $u$ transforms as a Lorentz scalar.

We now illustrate how the definitions of pairwise little group and pairwise helicity generalize to a state of two branes relevant to the calculation of the preceding section.  We then turn to multi-brane states in generality.

Consider an electric $(p - 1)$-brane and a magnetic $(d - p - 3)$-brane that span orthogonal directions in space, as in \eqref{coordinatelabels}.  In other words, we suppose that there exists a frame in which their spatial projection tensors $\Xi_1$ and $\Xi_2$ take the standard forms
\begin{align}
(\tilde{\Xi}_1)^{\mu\nu} &= -\operatorname{diag}(0, 0, 0, 0, \underbrace{1, \ldots, 1}_{p - 1}, \underbrace{0, \ldots, 0}_{d - p - 3}), \label{spatialproj1} \\
(\tilde{\Xi}_2)^{\mu\nu} &= -\operatorname{diag}(0, 0, 0, 0, \underbrace{0, \ldots, 0}_{p - 1}, \underbrace{1, \ldots, 1}_{d - p - 3}). \label{spatialproj2}
\end{align}
Suppose also that the branes have generic (spatially transverse) $d$-velocities $u_1$, $u_2$.  Via an $SO(3, 1)\subset SO(d - 1, 1)$ Lorentz transformation acting only on $(x_0, \vec{x})$, which preserves the spatial projection tensors \eqref{spatialproj1}--\eqref{spatialproj2}, any $u_1$, $u_2$ may be brought to the standard forms
\begin{align}
\tilde{u}_1^\mu = (u_1^0, 0, 0, +u_c, \vec{0}^{(p - 1)}, \vec{u}_1^{(d - p - 3)}), \label{refu1} \\
\tilde{u}_2^\mu = (u_2^0, 0, 0, -u_c, \vec{u}_2^{(p - 1)}, \vec{0}^{(d - p - 3)}), \label{refu2}
\end{align}
where $(u_i^0)^2 - u_c^2 - (\vec{u}_i)^2 = 1$ for $i = 1, 2$.  This frame makes manifest that there exists an $SO(2)$ subgroup of the Lorentz group that preserves $\tilde{\Xi}_1$, $\tilde{\Xi}_2$, and any two generic transverse $d$-velocities simultaneously.\footnote{Although the full pairwise little group of two branes in the configuration \eqref{spatialproj1}--\eqref{refu2} is $SO(2)\times SO(p - 2)\times SO(d - p - 4)$, we focus on the $SO(2)$ subgroup of external rotations because it will turn out to admit a transparent physical interpretation.}  When the $\vec{x}$-components of their velocities are equal and opposite along the $x_3$-axis as in \eqref{refu1}--\eqref{refu2}, the $SO(2)$ pairwise little group of these two branes consists of rotations $R_{12}$ in the plane of $x_1, x_2$.

We can now define a two-brane representation of the Lorentz group with respect to the reference projectors $(\tilde{\Pi}_i)^{\mu\nu} = (\tilde{u}_i)^\mu(\tilde{u}_i)^\nu + (\tilde{\Xi}_i)^{\mu\nu}$ defined by \eqref{spatialproj1}--\eqref{refu2}.  This representation is carried by states $|\Pi_1, \Pi_2; \sigma_1, \sigma_2; \sigma_{12}\rangle$, where the $\sigma_i$ label single-brane little group representations and $\sigma_{12}\equiv q_{12}$ labels a representation of the aforementioned $SO(2)$ pairwise little group.  These states transform according to
\begin{align}
&U(\Lambda)|\Pi_1, \Pi_2; \sigma_1, \sigma_2; q_{12}\rangle = \\
&e^{iq_{12}\phi_{12}}D_{\sigma_1'\sigma_1}D_{\sigma_2'\sigma_2}|\Lambda\Pi_1\Lambda^{-1}, \Lambda\Pi_2\Lambda^{-1}; \sigma_1', \sigma_2'; q_{12}\rangle, \nonumber
\end{align}
where the $D$'s are unitary representations of single-brane little group transformations.  As any pairwise little group transformation takes the form $W_{12} = R_{12}(\phi_{12})$ for some angle $\phi_{12} = \phi(\Lambda, \Pi_1, \Pi_2)$, we have written $D(W_{12}) = e^{iq_{12}\phi_{12}}$.  The $SO(2)$ charge $q_{12}$, or the pairwise helicity, can thus be identified with the electromagnetic angular momentum of these mutually nonlocal branes (which are asymptotically separated along the $x_3$-axis).  Furthermore, we have seen that $q_{12}$ satisfies a quantization condition that gives rise to a bona fide representation of $SO(2)$.

\section{Multi-Brane Representations of the Lorentz Group}

Finally, we define general states of asymptotically separated, non-intersecting branes in $\mathbb{R}^{d - 1, 1}$, in the spirit of the four-dimensional multiparticle states of \cite{Csaki:2020inw, Csaki:2020yei}. Such states may have quantum numbers that are not captured by tensor products of single-brane states.

We define a general $N$-brane Lorentz representation by a set of reference projection tensors $\{\tilde{\Pi}_i \,|\, i = 1, \ldots, N\}$ and a set of representations $\{\sigma_\ell \,|\, \ell\in \mathcal{L}\}$, where $\mathcal{L}$ indexes all possible little groups $G_\ell$, which are subgroups of the Lorentz group that stabilize the reference data. Explicitly, we take $\mathcal{L}$ to consist of all $2^N - 1$ nontrivial subsets of $\{1, \ldots, N\}$. For each $\ell\in \mathcal{L}$, we take $G_\ell$ to consist of those Lorentz transformations $\lambda\in SO(d - 1, 1)$ for which
\begin{equation}
\lambda\tilde{\Pi}_i\lambda^{-1} = \tilde{\Pi}_i \quad \text{for all $i\in \ell$}.
\end{equation}
Every state in this representation is defined in terms of some canonical Lorentz transformations relating it to the reference state $|\tilde{\Pi}_1, \ldots, \tilde{\Pi}_N; \{\sigma_\ell \,|\, \ell\in \mathcal{L}\}\rangle$, and an arbitrary state transforms as
\begin{align}
&U(\Lambda)|\Pi_1, \ldots, \Pi_N; \{\sigma_\ell \,|\, \ell\in \mathcal{L}\}\rangle = \\
&\left[\prod_{\ell\in \mathcal{L}} D_{\sigma_\ell'\sigma_\ell}(W_\ell)\right]|\Lambda\Pi_1\Lambda^{-1}, \ldots, \Lambda\Pi_N\Lambda^{-1}; \{\sigma_\ell' \,|\, \ell\in \mathcal{L}\}\rangle, \nonumber
\end{align}
where each $W_\ell\in G_\ell$ is a little group transformation that depends on $\{\Pi_i \,|\, i\in \ell\}$ as well as on $\Lambda$. Those $N$-brane representations that are induced by specifying nontrivial representations $\sigma_\ell$ only for $|\ell| = 1$ are tensor products of single-brane representations. In principle, however, we are free to specify a nontrivial representation of the entire little group $\prod_{\ell\in \mathcal{L}} G_\ell$. We have shown that the induced representations arising from nontrivial $\sigma_\ell$ for $|\ell| = 2$ have physical relevance, as was known to be the case even for particles \cite{Csaki:2020yei}.

In $d = 4$, all higher little groups beyond the pairwise ones are generically trivial. This is no longer true in higher dimensions. Therefore, it is natural to ask whether the higher little groups with $|\ell| > 2$ have any significance in $d > 4$. We content ourselves with addressing a simpler question: given a set of reference data for $n$ branes, what is the $n$-brane little group (for any $n = |\ell|$)?

To approach this question, it is simplest to begin with the case of particles. We define the $n$-particle little group as the subgroup of the $d$-dimensional Lorentz group that leaves invariant an arbitrary set of $n$ momenta (or velocities). For generic momenta, this group is $SO(d - n)$. The reasoning is simple: given $n$ generic (i.e., linearly independent) $d$-vectors, this $SO(d - n)$ preserves the orthogonal complement of their span.

Now consider $n$ branes with projection tensors $\Pi_i$ for $i = 1, \ldots, n$. Let $V_i$ denote the corresponding linear subspaces of Minkowski space $\mathbb{R}^{d - 1, 1}$, i.e., the images of the $\Pi_i$. Lorentz transformations that preserve a given subspace of Minkowski space act purely within the subspace or purely within its orthogonal complement. Therefore, those Lorentz transformations acting within any of the subspaces
\begin{equation}
V_1^{i_1}\cap\cdots \cap V_n^{i_n}
\label{intersection}
\end{equation}
for $(i_1, \ldots, i_n)\in \{0, 1\}^n$, where $V_i^0\equiv V_i$ and $V_i^1\equiv V_i^\perp$, preserve all subspaces $V_i$ simultaneously. The $n$-brane little group is thus at least as large as
\begin{equation}
\prod_{(i_1, \ldots, i_n)\in \{0, 1\}^n} \hspace{-1.5em} SO(\dim(V_1^{i_1}\cap\cdots \cap V_n^{i_n})),
\label{nbranelittlegroup}
\end{equation}
where the group $SO(m)$ is understood to be trivial for integers $m\leq 1$.\footnote{Note that the subspace \eqref{intersection} can contain a timelike direction only when $(i_1, \ldots, i_n) = (0, \ldots, 0)$. If it does, then the signature of the corresponding $SO$ factor in \eqref{nbranelittlegroup} should be adjusted accordingly. For our branes of interest, we would expect the corresponding $SO$ factor to be represented trivially, since it consists only of Lorentz transformations internal to all branes.} The actual little group may in fact be larger (in particular, we have ignored possible discrete subgroups).

Let us see what the formula \eqref{nbranelittlegroup} gives in special cases. If all $n$ branes are particles with generic $d$-momenta, then none of the momenta are collinear. Hence the only factor that contributes to \eqref{nbranelittlegroup} is that with $(i_1, \ldots, i_n) = (1, \ldots, 1)$, giving back the $n$-particle little group
\begin{equation}
SO(\dim(V_1^\perp\cap\cdots \cap V_n^\perp)) = SO(d - n),
\end{equation}
where we have used the fact that the intersection of $n$ generic hyperplanes ($(d - 1)$-dimensional subspaces) has dimension $d - n$. In the other extreme case that all $n$ branes are space-filling, we get simply
\begin{equation}
SO(\dim(V_1\cap\cdots \cap V_n) - 1, 1) = SO(d - 1, 1).
\end{equation}
For a single $(p - 1)$-brane ($n = 1$) spanning the subspace $V$, we get the expected little group
\begin{equation}
SO(\underbrace{\dim V - 1}_{p - 1}, 1)\times SO(\underbrace{\dim V^\perp}_{d - p}),
\end{equation}
including the internal Lorentz transformations that we take to be represented trivially. Finally, consider a $(p_1 - 1)$-brane and a $(p_2 - 1)$-brane. Generically, the intersection of $d_1$- and $d_2$-dimensional linear subspaces has dimension $\max(d_1 + d_2 - d, 0)$. So for generic configurations of these two branes, the corresponding pairwise little group always contains an $SO(d - p_1 - p_2)$ factor.\footnote{Note that the specific two-brane configuration in \eqref{spatialproj1}--\eqref{refu2} was chosen to simplify the angular momentum analysis for $p_1 = d \linebreak[1] - \linebreak[1] p_2 \linebreak[1] - 2\equiv p$, so the corresponding pairwise little group is larger than would be expected generically.}

We have not presented an exhaustive classification of multi-brane representations of the Lorentz group. While the subgroup of $SO(d - 1, 1)$ that preserves any given set of brane subspaces always contains a subgroup isomorphic to \eqref{nbranelittlegroup}, we have not attempted to find the full stabilizer subgroup or to catalogue all such groups that could arise. In addition, we have excluded non-generic ``tensionless'' brane representations with null (lightlike) $d$-velocities.

\section{Discussion}

In this paper, we have constructed representations of the Lorentz group that describe ideal branes whose worldvolume directions span linear subspaces of Minkowski spacetime. We have illustrated how quantum numbers associated to pairwise little groups of multi-brane representations can arise from the coupling of branes to abelian gauge fields in higher dimensions.

This study raises many more questions than answers.  We highlight some of these questions below.

\subsection{Origin of Brane Representations}

The representations of the Lorentz group constructed herein should be regarded as ``phenomenological'' models of a highly idealized class of brane states.  While these representations allow one to proceed by formal analogy to the construction of multiparticle states, their physical status remains unclear.  For instance, how might these brane representations embed into representations of the Poincar\'e group (or perhaps a different extension of the Lorentz group), and how might they be realized as physical states in a Poincar\'e-invariant theory of dynamical branes?

Unlike for particles, the usual Poincar\'e charges (energy and momentum) diverge in states of infinite branes, and are difficult to interpret as meaningful quantum numbers.  Said differently, because states of infinite branes occupy different superselection sectors than particle states, it is not at all obvious that a Hilbert space of branes should carry a representation of the Poincar\'e group, even in a theory that is classically Poincar\'e-invariant.\footnote{It should be kept in mind that this paper does not treat branes as defects with fixed asymptotic boundary conditions that spontaneously break $d$-dimensional Poincar\'e symmetry (such as flat defects $\mathbb{R}^{p - 1, 1}\subset \mathbb{R}^{d - 1, 1}$ that preserve $ISO(p - 1, 1)\times SO(d - p)\subset ISO(d - 1, 1)$), but rather as objects whose orientation in spacetime changes under Lorentz transformations.}  It may make more sense to ask whether there exists a local version of the Poincar\'e algebra, generated by currents rather than charges, that describes the Lorentz transformation properties of isotropic branes.  For example, the brane states defined here carry representations not only of the Lorentz algebra, but of an extension of the Lorentz algebra by two-index tensor generators.\footnote{The Poincar\'e group is a group extension of the Lorentz group by a vector representation thereof, while the states that we have defined form a representation of a group extension of the Lorentz group by a symmetric two-index tensor representation thereof.  The corresponding Lie algebra is
\begin{align*}
[\Pi_{\mu\nu}, \Pi_{\rho\sigma}] &= 0, \\
[M_{\mu\nu}, \Pi_{\rho\sigma}] &= i(\eta_{\mu\rho}\Pi_{\nu\sigma} - \eta_{\nu\rho}\Pi_{\mu\sigma}) + (\rho\leftrightarrow\sigma), \\
[M_{\mu\nu}, M_{\rho\sigma}] &= i(\eta_{\mu\rho}M_{\nu\sigma} - \eta_{\nu\rho}M_{\mu\sigma}) - (\rho\leftrightarrow\sigma).
\end{align*} \label{algebra}}  These extra generators carry two Lorentz indices regardless of the dimensionality of the brane, unlike the brane charges that occur in supersymmetry algebras \cite{deAzcarraga:1989mza, Dumitrescu:2011iu}.  They take finite values in states of infinite branes, but unlike currents, they have no position dependence.  Such an extended Lorentz algebra could conceivably be interpreted as a partially integrated version of the commutator algebra of the energy-momentum tensor $T^{\mu\nu}$ \cite{Deser:1967zzf}.

Alternatively, to directly assess the physical content of the abstract states $|\Pi^{\mu\nu}\rangle$, one could ask whether the geometrical data contained in $\Pi^{\mu\nu}$ admit an interpreta\-tion in terms of a maximal set of canonical variables that specify the state of an unexcited brane \cite{Regge:2016gaw}.

\subsection{Relevance to Brane Dynamics}

For our discussion, it is useful to make a distinction between ``kinematics'' (the definition of certain representations of the Lorentz group independently of any physical theory whose states might give rise to them) and ``dynamics'' (the interpretation of certain quantum numbers that label these representations in terms of parameters appearing in a Lagrangian).

Our treatments of brane kinematics and brane dynamics have been largely disjoint, and strengthening their connection remains an outstanding task.  On the dynamical side, we have presented a calculation of electromagnetic angular momentum for arbitrary branes related by electric-magnetic duality, generalizing earlier calculations for dyonic branes \cite{Deser:1997se}.  This calculation assumes a static configuration that suffices for deriving the form of the pairwise helicity.  On the kinematical side, we have presented a minimal mathematical model of how (not necessarily static) flat brane sources transform under the subgroups of the Lorentz group that realize these pairwise helicities as well as higher quantum numbers.  We have left open the question of whether this formalism has applications to dynamical questions about branes, and in particular, whether it provides a useful or meaningful way to describe brane states that are analogous to asymptotic states of particles.\footnote{The answer to this question may very well be ``no.''  For example, if the algebra of Footnote \ref{algebra} indeed arises as a partially integrated current algebra, then one would not expect the resulting $\Pi^{\mu\nu}$ to be conserved charges in the same sense as the Poincar\'e generators $P^\mu$.  In particular, one would not expect these quantities to be conserved in interactions where branes collide and fluctuate.}  For instance, could the pairwise quantum numbers of asymptotic brane states impose constraints on the scattering of electrically and magnetically charged branes \cite{Bachas:1995kx, McAllister:2004gd, DAmico:2014ywj}? (Some related comments appear in \cite{Csaki:2022ebw}.)

\subsection{Higher Little Groups for Particles and Branes}

In closing, we reiterate that while higher little groups are kinematically allowed in $d > 4$, it would be fascinating to identify (or to rule out) any dynamical effects that might lead to the $n$-particle or $n$-brane little group being represented nontrivially for $n > 2$.

\section{Acknowledgements}

I thank Csaba Cs\'aki, Ofri Telem, and John Terning for initial collaboration. I thank Jacques Distler, Dan Freed, Vadim Kaplunovsky, Andreas Karch, Liam McAllister, and Aaron Zimmerman for helpful discussions. Finally, I thank Steven Weinberg \cite{Weinberg:2020nsn} and Jacques Distler \cite{Distler:2020fzr} for providing the intellectual impetus to contemplate rep\-resentations of the Lorentz group in higher dimensions. This work was supported in part by the NSF grant PHY-1914679.

\bibliographystyle{apsrev4-1}
\bibliography{\jobname}

\appendix



\section{Pairwise Little Group} \label{app:pairwise}

\subsection{Definition for Particles}

Wigner \cite{Wigner:1939cj} classified the unitary irreducible (or one-particle) representations of the Poincar\'e group\footnote{More precisely, these are projective representations of the inhomogeneous proper orthochronous Lorentz group \cite{Weinberg:1995mt, Bekaert:2006py}.} according to their mass and their little group representation. The little group is the subgroup of the Lorentz group that leaves a particular reference momentum $k$ invariant. The single-particle Hilbert space is spanned by momentum eigenstates $|p; \sigma\rangle$, where the single-particle quantum numbers $\sigma$ fix the little group transformations of $|k; \sigma\rangle$. General states are obtained from this reference state by $|p; \sigma\rangle\equiv U(L_p)|k; \sigma\rangle$, where $U(L_p)$ is a unitary representation of the Lorentz boost $L_p$ such that $p = L_p k$. Under a general boost $\Lambda$, the transformation of $|p; \sigma\rangle$ is induced from that of $|k; \sigma\rangle$ as follows:
\begin{equation}
U(\Lambda)|p; \sigma\rangle = U(L_{\Lambda p})U(W)|k; \sigma\rangle = D_{\sigma'\sigma}(W)|\Lambda p; \sigma'\rangle,
\end{equation}
where $W\equiv L_{\Lambda p}^{-1}\Lambda L_p$ is a little group transformation and $D$ is a unitary representation of $W$.

Multiparticle representations of the Poincar\'e group are not limited to tensor products of single-particle representations. They may involve additional quantum numbers that characterize the Lorentz transformations of multiparticle states relative to tensor products of one-particle states \cite{Csaki:2020inw, Csaki:2020yei}.

To see this, we specialize to 4D, where these less familiar multiparticle states find their natural home in the study of scattering amplitudes for electric and magnetic charges. For a single massive or massless particle, we choose $k = (m, 0, 0, 0)$ or $k = (E, 0, 0, E)$, respectively, in which case the little group is $SU(2)$ or $ISO(2)$. For a \emph{pair} of particles with momenta $(p_1, p_2)$, we can always\footnote{Except when both momenta are null and parallel, a case that we ignore.} boost to a center-of-momentum (COM) frame in which the momenta take the form
\begin{equation}
\tilde{k}_{1, 2} = \left(\sqrt{m_{1, 2}^2 + p_c^2}, 0, 0, \pm p_c\right),
\end{equation}
where $p_c$ is the Lorentz-invariant COM momentum. Regardless of whether the particles are massive or massless, there exists a $U(1)$ pairwise little group of rotations about the $z$-axis that preserves these reference momenta.\footnote{Again, we ignore non-generic configurations of momenta in which the pairwise little group is enhanced to $SO(3)$.} To apply Wigner's method to multiparticle states, we relate the momenta of any pair of particles to the above reference momenta via Lorentz transformations:
\begin{align}
p_i &= L^i_{p_i}k_i, \\
(p_1, p_2) &= \left(L^{12}_{p_1, p_2}\tilde{k}_1, L^{12}_{p_1, p_2}\tilde{k}_2\right),
\end{align}
where $i = 1, 2$. The condition that $L^{12}_{p_1, p_2}$ takes $\smash{\tilde{k}_1}\rightarrow p_1$ \emph{and} $\smash{\tilde{k}_2}\rightarrow p_2$ determines it up to a $U(1)$ rotation. We then define the single-particle and pairwise little group transformations
\begin{align}
W_i &\equiv \left(L^{i}_{\Lambda p_i}\right)^{-1}\Lambda L^i_{p_i}, \\
W_{12} &\equiv \left(L^{12}_{\Lambda p_1, \Lambda p_2}\right)^{-1}\Lambda L^{12}_{p_1, p_2},
\end{align}
where the latter is a rotation: $R_z(\phi_{12})\equiv W_{12}$. Finally, since three-particle and higher little groups are trivial in 4D, a general $n$-particle state takes the form
\begin{equation}
|p_1, \ldots, p_n; \sigma_1, \ldots, \sigma_n; q_{12}, \ldots, q_{n-1, n}\rangle,
\end{equation}
where $p_i$ is the momentum of particle $i$, the $\sigma_i$ are single-particle quantum numbers (spins or helicities), and there are $\binom{n}{2}$ quantized pairwise helicities, or $U(1)$ charges, $q_{ij}$ associated with pairs of particles $i, j$. Under a Lorentz transformation $\Lambda$, it transforms as
\begin{align}
&U(\Lambda)|p_1, \ldots, p_n; \sigma_1, \ldots, \sigma_n; q_{12}, \ldots, q_{n-1, n}\rangle = \prod_{i<j} e^{iq_{ij}\phi_{ij}} \nonumber \\
&\times \prod_{i=1}^n D_{\sigma_i'\sigma_i}^i|\Lambda p_1, \ldots, \Lambda p_n; \sigma_1', \ldots, \sigma_n'; q_{12}, \ldots, q_{n-1, n}\rangle.
\end{align}
The rotation angles $\phi_{ij} = \phi(\Lambda, p_i, p_j)$ parametrize pairwise little group transformations, while the $D$'s are unitary representations of one-particle little group transformations: $D_{\sigma_i'\sigma_i}^i = D_{\sigma_i'\sigma_i}(W_i)$.\footnote{Note that using rest- or COM-frame reference momenta is not essential to this construction. Alternatively, one could specify a set of $n$ reference momenta $\tilde{p}_1, \ldots, \tilde{p}_n$ at the outset, which fix all one-particle and pairwise little groups.}

The pairwise helicities $q_{ij}$ depend on the dynamics of the theory. If particles $i, j$ are dyons, then $q_{ij}$ is simply the quantized angular momentum $\frac{1}{4\pi}(e_i g_j - e_j g_i)$ in their electromagnetic field.

\subsection{Definition for Branes}

We now turn to branes in any dimension. We define $(p - 1)$-brane states that, for $p > 1$, yield representations of the Lorentz group but not of the full Poincar\'e group, as they are not obtained by diagonalizing the translation generators of the Poincar\'e algebra.

The state of a single brane at rest that extends along the first $p - 1$ spatial axes is written as $|\tilde{\Pi}; \sigma\rangle$, with the reference projection tensor $\tilde{\Pi}$ given in \eqref{referencepi}. A little group rotation $W$ acts on this reference state as
\begin{equation}
U(W)|\tilde{\Pi}; \sigma\rangle = D_{\sigma'\sigma}(W)|\tilde{\Pi}; \sigma'\rangle,
\end{equation}
where the $D$'s are matrix elements of unitary representations of $W$. A general spacetime orientation $\Pi$ is related to \eqref{referencepi} by a Lorentz transformation, which we denote by $L(\Pi)$:
\begin{equation}
\Pi = L\tilde{\Pi}L^{-1}.
\end{equation}
A generic brane state is defined by
\begin{equation}
|\Pi; \sigma\rangle = U(L)|\tilde{\Pi}; \sigma\rangle.
\end{equation}
Finally, a general Lorentz transformation $\Lambda$ acts as
\begin{equation}
U(\Lambda)|\Pi; \sigma\rangle = D_{\sigma'\sigma}(W)|\Lambda\Pi\Lambda^{-1}; \sigma'\rangle,
\end{equation}
where the little group rotation $W$ is given by
\begin{equation}
W(\Lambda, \Pi) = L^{-1}(\Lambda\Pi\Lambda^{-1})\Lambda L(\Pi).
\end{equation}
The single-brane states $|\Pi; \sigma\rangle$ furnish an infinite-di\-men\-sion\-al unitary representation of the $d$-dimensional Lorentz group.

We define a two-brane representation of the Lorentz group by specifying a reference state
\begin{equation}
|\tilde{\Pi}_1, \tilde{\Pi}_2; \sigma_1, \sigma_2; \sigma_{12}\rangle
\label{twobranereference}
\end{equation}
for some reference projectors $\tilde{\Pi}_i$, single-brane little group representations $\sigma_i$, and pairwise little group representation $\sigma_{12}$. Given these data, we define a general state in this representation by its transformation property
\begin{align}
U&(\Lambda)|\Pi_1, \Pi_2; \sigma_1, \sigma_2; \sigma_{12}\rangle \nonumber \\
&= D_{\sigma_1'\sigma_1}(W_1)D_{\sigma_2'\sigma_2}(W_2)D_{\sigma_{12}'\sigma_{12}}(W_{12}) \label{twobranetransformation} \\
&\phantom{==} \times |\Lambda\Pi_1\Lambda^{-1}, \Lambda\Pi_2\Lambda^{-1}; \sigma_1', \sigma_2'; \sigma_{12}'\rangle, \nonumber
\end{align}
where
\begin{align}
W_i &= L_i(\Lambda\Pi_i\Lambda^{-1})^{-1}\Lambda L_i(\Pi_i), \\
W_{12} &= L_{12}(\Lambda\Pi_1\Lambda^{-1}, \Lambda\Pi_2\Lambda^{-1})^{-1}\Lambda L_{12}(\Pi_1, \Pi_2)
\end{align}
are (pairwise) little group transformations and $L_i(\Pi_i)$, $L_{12}(\Pi_1, \Pi_2)$ are some standard Lorentz transformations satisfying
\begin{equation}
\Pi_i = L_i\tilde{\Pi}_i L_i^{-1}, \qquad \Pi_i = L_{12}\tilde{\Pi}_i L_{12}^{-1}
\end{equation}
for $i = 1, 2$.

In more detail, a two-brane reference state of the form \eqref{twobranereference} should be thought of as a tensor product
\begin{equation}
|\tilde{\Pi}_1; \sigma_1\rangle\otimes |\tilde{\Pi}_2; \sigma_2\rangle\otimes |(\tilde{\Pi}_1, \tilde{\Pi}_2); \sigma_{12}\rangle, \label{twobranereferencetensor}
\end{equation}
where the first two factors in \eqref{twobranereferencetensor} are single-brane reference states and the last factor defines a family of \emph{pairwise} states, which are auxiliary objects that transform analogously to single-brane states under the Lorentz group, but with respect to Lorentz transformations that act simultaneously on a pair of brane projectors $(\Pi_1, \Pi_2)$. Given \eqref{twobranereferencetensor}, we define a \emph{generalized} two-brane state as
\begin{align}
&|\Pi_1, \Pi_2, (\Theta_1, \Theta_2); \sigma_1, \sigma_2; \sigma_{12}\rangle \\
&\equiv \left(\bigotimes_{i=1}^2 U(L_i)|\tilde{\Pi}_i; \sigma_i\rangle\right)\otimes U(L_{12})|(\tilde{\Pi}_1, \tilde{\Pi}_2); \sigma_{12}\rangle, \nonumber
\end{align}
where $L_i(\Pi_i)$ and $L_{12}(\Theta_1, \Theta_2)$ are some standard Lorentz transformations satisfying (for $i = 1, 2$)
\begin{equation}
\Pi_i = L_i\tilde{\Pi}_i L_i^{-1}, \qquad \Theta_i = L_{12}\tilde{\Pi}_i L_{12}^{-1}.
\end{equation}
Note that while each $L_i$ takes $\tilde{\Pi}_i$ to $\Pi_i$ individually, $L_{12}$ takes the pair $(\tilde{\Pi}_1, \tilde{\Pi}_2)$ to $(\Theta_1, \Theta_2)$. The definitions of the $L_i$ and of $L_{12}$ are unique up to single-brane and pairwise little group transformations, respectively. An arbitrary Lorentz transformation $\Lambda$ therefore acts on generalized two-brane states as
\begin{align}
&U(\Lambda)|\Pi_1, \Pi_2, (\Theta_1, \Theta_2); \sigma_1, \sigma_2; \sigma_{12}\rangle \nonumber \\
&= \left(\bigotimes_{i=1}^2 U(W_i)|\Lambda\Pi_i\Lambda^{-1}; \sigma_i\rangle\right) \\
&\phantom{==} \otimes U(W_{12})|(\Lambda\Theta_1\Lambda^{-1}, \Lambda\Theta_2\Lambda^{-1}); \sigma_{12}\rangle \nonumber \\
&= \left(\bigotimes_{i=1}^2 D_{\sigma_i'\sigma_i}(W_i)|\Lambda\Pi_i\Lambda^{-1}; \sigma_i'\rangle\right) \\
&\phantom{==} \otimes D_{\sigma_{12}'\sigma_{12}}(W_{12})|(\Lambda\Theta_1\Lambda^{-1}, \Lambda\Theta_2\Lambda^{-1}); \sigma_{12}'\rangle, \nonumber
\end{align}
where
\begin{align}
W_i &= L_i(\Lambda\Pi_i\Lambda^{-1})^{-1}\Lambda L_i(\Pi_i), \label{Wi} \\
W_{12} &= L_{12}(\Lambda\Theta_1\Lambda^{-1}, \Lambda\Theta_2\Lambda^{-1})^{-1}\Lambda L_{12}(\Theta_1, \Theta_2) \label{W12}
\end{align}
are (pairwise) little group transformations. Finally, we define a \emph{physical} two-brane state as one satisfying $\Theta_i = \Pi_i$ for $i = 1, 2$:
\begin{align}
&|\Pi_1, \Pi_2; \sigma_1, \sigma_2; \sigma_{12}\rangle \nonumber \\
&\equiv |\Pi_1, \Pi_2, (\Pi_1, \Pi_2); \sigma_1, \sigma_2; \sigma_{12}\rangle.
\end{align}
Such states transform as written in \eqref{twobranetransformation}, with $W_i$ and $W_{12}$ as in \eqref{Wi}--\eqref{W12} but with $\Theta_i = \Pi_i$. The collection of generalized two-brane states defines a unitary representation of the Lorentz group, and that of physical two-brane states defines a subrepresentation.

\subsection{Hilbert Space Interpretation}

We now make some overdue comments on the Hilbert space structure of the representations that we have defined, and in particular, on the inner products with respect to which they are unitary.

Taking a broad perspective, for a classical system with abstract configuration space $X$, we are often interested in the corresponding quantum system with Hilbert space $L^2(X)$.\footnote{More generally, we may wish to consider the space of $L^2$ sections of a complex vector bundle over $X$, where the fiber $V$ parametrizes internal degrees of freedom.  For a trivial bundle, the Hilbert space is simply $L^2(X)\otimes V$.}  We can write the inner product on $L^2(X)$ as
\begin{equation}
\langle f_1|f_2\rangle = \int dx\, f_1(x)^\ast f_2(x),
\end{equation}
where $f_1, f_2 : X\to \mathbb{C}$ and $x\in X$.  If the group $G$ acts on $X$ and the associated measure $dx$ is $G$-invariant, then $L^2(X)$ carries a unitary representation $U$ of $G$ induced by the action of $G$ on $X$.  For $g\in G$, we have
\begin{equation}
U(g)|f\rangle = |f^g\rangle
\end{equation}
where $f^g(x) = f(g^{-1}x)$.  In terms of ``position eigenstates'' that satisfy $\langle x|x'\rangle = \delta(x - x')$ with respect to the measure $dx$ (and are therefore not elements of $L^2(X)$), this action is equivalent to
\begin{equation}
U(g)|x\rangle = |gx\rangle,
\end{equation}
where $|f\rangle = \int dx\, f(x)|x\rangle$.

As a special case of the above, the carrier space of a unitary irreducible representation of the Poincar\'e group corresponding to a spinless particle of mass $m > 0$ is the space of $L^2$ functions on the positive mass shell $S_m^+$, i.e., the orbit of the fiducial timelike $d$-vector $k^\mu = (m, 0, \ldots, 0)$ in $\mathbb{R}^{d - 1, 1}$ under proper orthochronous Lorentz transformations.  Including spin, the carrier space is the space of $L^2$ sections of a vector bundle with fiber $V$ over $S_m^+$, where $V$ is the carrier space of a unitary irreducible representation of the little group $SO(d - 1)$.  $S_m^+$ inherits a Lorentz-invariant integration measure from the flat measure on $\mathbb{R}^{d - 1, 1}$ via the embedding $S_m^+\subset \mathbb{R}^{d - 1, 1}$, which we use to define an inner product:
\begin{equation}
\langle\Psi|\Phi\rangle = \int d^d p\, \delta(p^2 - m^2)\theta(p^0)\sum_\sigma \Psi_{p, \sigma}^\ast\Phi_{p, \sigma}.
\end{equation}
Lorentz transformations act on states as $U(\Lambda)|\Psi\rangle = |\Psi^\Lambda\rangle$ with
\begin{equation}
\Psi_{p, \sigma}^\Lambda = \sum_{\sigma'} D_{\sigma\sigma'}(W^{-1}(\Lambda^{-1}, p))\Psi_{\Lambda^{-1}p, \sigma'},
\end{equation}
where $W$ is a little group transformation with respect to $k^\mu$ and we have used $W(\Lambda, \Lambda^{-1}p) = W^{-1}(\Lambda^{-1}, p)$.  We write the wavefunctions $\Psi_{p, \sigma}$ in terms of $d$-momenta $p^\mu$ rather than $(d - 1)$-momenta $\vec{p}$, with the understanding that the $d$-momenta are constrained to lie on $S_m^+$.

Similarly, ignoring spin, the carrier space of a single-brane representation of the Lorentz group as defined in this paper is the space of $L^2$ functions on the Lorentz orbit of a reference projection tensor within the space of all two-index Lorentz tensors.  Such an orbit inherits a Lorentz-invariant integration measure from the ambient space, and this measure allows one to define an inner product on the corresponding Hilbert space.

More explicitly, Lorentz transformations $\Lambda$ act as $\Pi\to \Lambda\Pi\Lambda^{-1}$ on the set of all two-index Lorentz tensors $\Pi$ (thought of as matrices), which is isomorphic to $\mathbb{R}^{d^2}$ as a vector space.  This space has a natural Lorentz-invariant flat measure, which we denote by $d\Pi$.\footnote{Whereas $d^d p$ transforms with a factor of $|{\det\Lambda}|$, $d\Pi$ transforms with a factor of $|{\det\Lambda}|^{2d}$.  This is easily seen by considering the vectorization of $\Pi$ and the behavior of the determinant under Kronecker products.}  Let $O_p$ denote the orbit of the fiducial rank-$p$ projection tensor
\begin{equation}
\tilde{\Pi} = \operatorname{diag}(\underbrace{1, \ldots, 1}_{p}, \underbrace{0, \ldots, 0}_{d - p})
\label{fiducial}
\end{equation}
under the action of the Lorentz group, and let $d\Pi_p$ denote the measure on $O_p$ that is induced by the flat measure $d\Pi$ on the space of all $\Pi$.  Then we obtain an infinite-dimensional unitary representation of the Lorentz group on $L^2(O_p)$ given (in terms of ``position eigenstates'') by $U(\Lambda)|\Pi\rangle = |\Lambda\Pi\Lambda^{-1}\rangle$, or equivalently (in terms of normalizable states) by $U(\Lambda)|\Psi\rangle = |\Psi^\Lambda\rangle$ with $\Psi_\Pi^\Lambda = \Psi_{\Lambda^{-1}\Pi\Lambda}$.  We have written the wavefunctions as functions of $\Pi$, with $\Pi$ implicitly constrained to $O_p$.  One can also add spin indices, using the fact that \eqref{fiducial} has a nontrivial stabilizer within the Lorentz group, in which case the carrier space is enlarged and the inner product takes the form
\begin{equation}
\langle\Psi|\Phi\rangle = \int d\Pi_p\sum_\sigma \Psi_{\Pi, \sigma}^\ast\Phi_{\Pi, \sigma}.
\end{equation}
In general, we would expect such a representation to be reducible.

Finally, while massive particle representations of the Poincar\'e group and these unitary representations of the Lorentz group share the common feature of being induced by unitary representations of compact subgroups of the Lorentz group, these brane representations admit no natural action of the translation generators of the Poincar\'e group.  Indeed, the total energies and momenta of infinite branes do not provide meaningful quantum numbers for labeling states, while their finite energy and momentum densities transform non-tensorially (unlike the generators of the Poincar\'e algebra).

\section{Mathematical Conventions} \label{app:conventions}

We denote the dimension of spacetime by $d$.  We assume that spacetime is topologically trivial and boundaryless.  For our subsequent calculations (unlike in the main text), we work in ``mostly plus'' Lorentzian signature.  We define the Hodge star operator by
\begin{align}
&\ast(dx^{\mu_1}\wedge \cdots\wedge dx^{\mu_p}) \\
&= \frac{1}{(d - p)!}\epsilon^{\mu_1\cdots \mu_p}{}_{\nu_1\cdots \nu_{d-p}}dx^{\nu_1}\wedge \cdots\wedge dx^{\nu_{d-p}}. \nonumber
\end{align}
In particular, $dx^{\mu_1}\wedge \cdots\wedge dx^{\mu_d} = -\epsilon^{\mu_1\cdots \mu_d}\ast 1$ with $\ast 1 = d^d x\sqrt{-g}$.  With this convention, for a $p$-form
\begin{equation}
\alpha = \frac{1}{p!}\alpha_{\mu_1\cdots \mu_p}dx^{\mu_1}\wedge\cdots \wedge dx^{\mu_p},
\end{equation}
the components of its dual are given by
\begin{equation}
(\ast\alpha)_{\mu_{p+1}\cdots \mu_d} = \frac{1}{p!}\epsilon_{\mu_1\cdots \mu_d}\alpha^{\mu_1\cdots \mu_p}.
\end{equation}
For any two $p$-forms $\alpha$ and $\beta$, we have
\begin{equation}
\alpha\wedge \ast\beta = \beta\wedge \ast\alpha = \frac{1}{p!}\alpha_{\mu_1\cdots \mu_p}\beta^{\mu_1\cdots \mu_p}\ast 1,
\end{equation}
as $\epsilon_{\mu_1\cdots \mu_p\rho_1\cdots \rho_{d-p}}\epsilon^{\nu_1\cdots \nu_p\rho_1\cdots \rho_{d-p}} = -p!(d - p)!\delta_{\mu_1\cdots \mu_p}^{\nu_1\cdots \nu_p}$.

Poincar\'e duality associates (cohomology classes of) $p$-forms with (homology classes of) codimension-$p$ submanifolds \cite{Bott:1982}.  Operationally, given a $p$-dimensional submanifold $M$, its Poincar\'e dual $\smash{\widehat{M}}$ has delta-function support on $M$ and is defined as satisfying
\begin{equation}
\int_M A = \int A\wedge \widehat{M}
\end{equation}
for all $p$-forms $A$, where the integral on the right is taken over all of spacetime.  For a $p$-manifold $M$, we have
\begin{equation}
\widehat{\partial M} + (-1)^{d - p}d\widehat{M} = 0.
\label{morphism}
\end{equation}
Poincar\'e duality is an involution, unlike Hodge duality.

If $M$ and $N$ are submanifolds with $\dim M + \dim N = d$ that intersect transversally at a finite number of points, then the (signed) intersection number is an integer given by
\begin{equation}
I(M, N) = \int \widehat{M}\wedge \widehat{N},
\end{equation}
reflecting the fact that cup product in cohomology is Poincar\'e-dual to intersection pairing in homology.  It satisfies $I(M, N) = (-1)^{\dim M\dim N}I(N, M)$.

If $C$ and $D$ are two non-intersecting trivial cycles in spacetime with $\dim C + \dim D = d - 1$, where $C = \partial C'$ and $D = \partial D'$, then we define the linking number of $C$ and $D$ as the intersection number of $C$ and the ``Stokes surface'' $D'$:
\begin{equation}
L(C, D) = \int \widehat{C}\wedge \widehat{D'}.
\end{equation}
It satisfies $L(C, D) = (-1)^{(\dim C + 1)(\dim D + 1)}L(D, C)$ as a consequence of \eqref{morphism}.

\section{Pairwise Helicity From Angular Momentum} \label{app:angularmomentum}

In this appendix, we elaborate on the calculation of the angular momentum \eqref{theresult}.

We first recall the argument for the Dirac(-Schwinger-Zwanziger) quantization condition. Consider an electric brane with $p$-dimensional worldvolume $V$ and a spacelike submanifold $\Sigma^{d - p}$. The quantity
\begin{equation}
Q = \int_{\Sigma^{d - p}} \ast J = \int_{(\partial\Sigma)^{d - p - 1}} \ast F
\label{charge}
\end{equation}
defines a conserved charge in the sense that it depends only on the linking homology class of $(\partial\Sigma)^{d - p - 1}$ with respect to $V$. Here, we have applied Stokes' theorem to the electric brane current $J$: $d\ast F = \ast J$. Therefore, the charge densities of an electric $(p - 1)$-brane and a magnetic $(d - p - 3)$-brane, as measured by spheres $S^{d - p - 1}$ and $S^{p + 1}$ that link them in space, are given by
\begin{equation}
e = \int_{S^{d - p - 1}} \ast F, \qquad g = \int_{S^{p + 1}} F,
\end{equation}
where we assume that $1\leq p\leq d - 3$. More generally, \eqref{charge} evaluates to $Q = eI(V, \Sigma) = eL(V, \partial\Sigma)$, where $I$ and $L$ denote intersection and linking numbers, respectively.

Now note that in a background with nonzero magnetic charge, $A$ cannot be globally defined, since otherwise, we would have $F = dA$ and $g = 0$. Hence the action of an electric brane is ambiguous in such a background. If an electric brane sweeps out a closed Wilson surface along $\Sigma^p$ in the field of a magnetic brane, then the amplitude for this process is given by the exponentiated action
\begin{equation}
\exp\left(ie\int_{\Sigma^p} A\right) = \exp\left(ie\int_{M^{p+1}} F\right),
\end{equation}
where $M^{p+1}$ is any $(p + 1)$-manifold with boundary $\Sigma^p$. The expression on the right is unambiguous because $F$ is globally defined, but it depends nonlocally on $\Sigma^p$. To preserve locality, we demand that it be independent of the choice of $M^{p+1}$, or equivalently, that
\begin{equation}
\exp\left(ie\int_{C^{p+1}} F\right) = \exp(iegL(V', C^{p+1})) = 1
\end{equation}
for any closed $C^{p+1}$, where $V'$ is the worldvolume of the magnetic brane. In particular, choosing $C^{p+1}$ to have unit linking number with the magnetic brane in space yields $\exp(ieg) = 1$, or the Dirac quantization condition
\begin{equation}
eg\in 2\pi\mathbb{Z}.
\label{Dirac}
\end{equation}
This condition can be derived in many alternative ways, such as via Dirac branes \cite{Nepomechie:1984wu, Teitelboim:1985yc} or by considering the transition functions between gauge patches \cite{Henneaux:1986ht} (generalizing the Wu-Yang construction \cite{Wu:1975es}).

While \eqref{Dirac} holds for generic $d$ and $p$, in the special case that $d = 2(p + 1)$, brane sources can carry both electric and magnetic charge. In this case, given any two dyonic $(p - 1)$-branes of charge densities $(e_1, g_1)$ and $(e_2, g_2)$, we instead have the condition
\begin{equation}
e_1 g_2 + (-1)^p e_2 g_1\in 2\pi\mathbb{Z},
\label{DSZ}
\end{equation}
which we refer to as the DSZ quantization condition. The derivation of \eqref{DSZ} is more subtle than that of \eqref{Dirac} \cite{Bremer:1997qb, Deser:1997se, Bertolini:1998mg, Deser:1998vc, Bekaert:2002cz}. The presence of the sign $(-1)^p$ is closely related to the (non)existence of theta terms and the structure of the electric-magnetic duality group in the corresponding $d$.\footnote{The theta angle modifies the electric charges of dyons via the Witten effect \cite{Witten:1979ey} while preserving the DSZ pairing. Only the latter enters into pairwise helicities, not the individual charges. But scattering processes of dyonic particles or branes are sensitive to the charges themselves and therefore also to the theta angle.} See Appendix \ref{app:pform} for details.

The conditions \eqref{Dirac} and \eqref{DSZ} motivate the definition \eqref{defpairwise} for the pairwise helicity of two branes $i$ and $j$ related by electric-magnetic duality. We have shown that any two such branes admit a kinematic configuration in which their pairwise little group is $SO(2)$. We now show that \eqref{defpairwise} arises dynamically as the $SO(2)$ charge. Specifically, we show that the pairwise helicity $q_{ij}$ is the angular momentum in the electromagnetic field sourced by dual branes. This angular momentum shows up as a phase in the pairwise little group transformation of asymptotic brane states. As a byproduct of our analysis, we also (re)derive the conditions \eqref{Dirac} and \eqref{DSZ} in a uniform way.

\subsection{Pairwise Helicity for Particles}

Our derivation proceeds by analogy to the case $d = 4$, $p = 1$, which we now review. In 4D, particles (0-branes) can carry both electric and magnetic charge. The DSZ quantization condition \eqref{DSZ} becomes
\begin{equation}
q_{12}\equiv \frac{e_1 g_2 - e_2 g_1}{4\pi}\in \tfrac{1}{2}\mathbb{Z}.
\label{DSZ4D}
\end{equation}
We first recall how $q_{12}$ appears as the angular momentum of a charge-monopole field, setting $(e_1, g_1) = (0, g)$ and $(e_2, g_2) = (e, 0)$.  For ease of generalization, we work in terms of differential forms.

We place the monopole at the origin in $\vec{x} = (x_1, x_2, x_3)$ and the electric charge at $\smash{\vec{R}} = (R_1, R_2, R_3)$. We normalize their field strengths and potentials as follows:
\begin{equation}
\vec{E} = \frac{e(\vec{x} - \vec{R})}{4\pi|\vec{x} - \vec{R}|^3}, \qquad \vec{B} = \frac{g\vec{x}}{4\pi|\vec{x}|^3},
\end{equation}
where $E^i\equiv F^{0i}$ and $B^i\equiv -(\ast F)^{0i} = \frac{1}{2}\epsilon^{ijk}F_{jk}$. In terms of the one-forms $E = E_i\, dx^i$ and $B = B_i\, dx^i$, the angular momentum pseudovector ($J^i = \frac{1}{2}\epsilon^{ijk}J_{jk}$) becomes
\begin{equation}
J_{ij} = \int (x_i\, dx_j - x_j\, dx_i)\wedge E\wedge B.
\label{formula}
\end{equation}
Substituting $B = \frac{g}{4\pi|\vec{x}|^3}x_k\, dx^k$ into \eqref{formula} gives
\begin{equation}
J_{ij} = \frac{g}{4\pi}\int \frac{E}{|\vec{x}|^3}\wedge (|\vec{x}|^2\, dx_i\wedge dx_j - \epsilon_{ijk}x^k\ast' (x_\ell\, dx^\ell)),
\end{equation}
where we use $\ast'$ to denote the Hodge star with respect to the (Euclidean) spatial dimensions. This simplifies to
\begin{equation}
J_{ij} = \frac{g}{4\pi}\int E\wedge\ast' d\left(\frac{\epsilon_{ijk}x^k}{|\vec{x}|}\right) = -\frac{g}{4\pi}\int \frac{\epsilon_{ijk}x^k}{|\vec{x}|}\, d(\ast' E),
\end{equation}
where we have used symmetry of the inner product on differential forms and integrated by parts. Using $d(\ast' E) = (\nabla\cdot \vec{E})\, d^3 x = e\delta^3(\vec{x} - \vec{R})\, d^3 x$ then gives
\begin{equation}
J_{ij} = -\frac{eg}{4\pi}\epsilon_{ijk}\hat{R}^k
\label{4dresult}
\end{equation}
where $\hat{R}\equiv \vec{R}/|\vec{R}|$, which is independent of the separation $\smash{|\vec{R}|}$. The Dirac quantization condition now follows from the semiclassical expectation that the angular momentum \eqref{4dresult} should be half-integrally quantized.\footnote{To repeat this derivation in non-covariant language \cite{Shnir:2005vvi}, substituting $\vec{B}$ into $\vec{J} = \int d^3 x\, [\vec{x}\times (\vec{E}\times \vec{B})]$ gives
\begin{equation}
\vec{J} = \frac{g}{4\pi}\int d^3 x\, (\vec{E}\cdot \nabla)\hat{x} = -\frac{g}{4\pi}\int d^3 x\, (\nabla\cdot \vec{E})\hat{x} = -\frac{eg}{4\pi}\hat{R},
\end{equation}
where $\hat{x}\equiv \vec{x}/|\vec{x}|$ and we have dropped a surface term since the fields vanish at infinity.}

\subsection{Pairwise Helicity for Branes}

We now calculate the angular momentum in the electromagnetic field of two mutually nonlocal branes in $\mathbb{R}^{d - 1, 1}$, obtaining precisely the pairwise helicities \eqref{defpairwise}.

To start, the gauge field carries a momentum density
\begin{equation}
P_i = T^0{}_i = \frac{1}{p!}F^0{}_{i_1\cdots i_p}F_{i}{}^{i_1\cdots i_p},
\end{equation}
where $T_{\mu\nu} = -2\frac{\partial\mathcal{L}}{\partial g^{\mu\nu}} + g_{\mu\nu}\mathcal{L}$ is the stress tensor of the $p$-form Maxwell theory with Lagrangian density
\begin{equation}
\mathcal{L} = -\frac{1}{2(p + 1)!}F_{\mu_1\cdots \mu_{p+1}}F^{\mu_1\cdots \mu_{p+1}}.
\end{equation}
The corresponding angular momentum density is
\begin{equation}
\mathcal{J}_{ij} = x_i P_j - x_j P_i = x_i T^0{}_j - x_j T^0{}_i,
\end{equation}
which is a rank-two antisymmetric tensor of $SO(d - 1)$. The angular momentum $J_{ij} = \int d^{d-1} x\, \mathcal{J}_{ij}$ is dimensionless in natural units.

We write the electric and magnetic fields as (spatial) forms $E$ and $B$, with components $E^{i_1\cdots i_p} = F^{0i_1\cdots i_p}$ and
\begin{align}
B^{i_1\cdots i_{d-p-2}} &= (-1)^p(\ast F)^{0i_1\cdots i_{d-p-2}} \\
&= \frac{1}{(p + 1)!}\epsilon^{j_1\cdots j_{p+1}i_1\cdots i_{d-p-2}}F_{j_1\cdots j_{p+1}}.
\end{align}
Note that $\epsilon_{0i_1\cdots i_{d-1}} = \epsilon_{i_1\cdots i_{d-1}}$ and $\epsilon^{0i_1\cdots i_{d-1}} = -\epsilon^{i_1\cdots i_{d-1}}$. We denote the spatial Hodge star by $\ast'$, so that
\begin{equation}
dx^{i_1}\wedge\cdots \wedge dx^{i_{d-1}} = \epsilon^{i_1\cdots i_{d-1}}\ast' 1.
\end{equation}
We then compute that
\begin{equation}
(x_i\, dx_j - x_j\, dx_i)\wedge E\wedge B = (x_i T^0{}_j - x_j T^0{}_i)\ast' 1.
\label{integrand}
\end{equation}
It follows from \eqref{integrand} that the formula \eqref{formula} for the angular momentum corresponding to $SO(2)$ rotations in the $i, j$ directions holds in arbitrary $d$.

We restrict our attention to the spatial directions $\mathbb{R}^{d - 1}$. We take the electric $(p - 1)$-brane to fill the directions
\begin{equation}
\vec{R}_3\times \mathbb{R}^{p - 1}\times \vec{0}_{d - p - 3}
\end{equation}
and the magnetic $(d - p - 3)$-brane to fill the directions
\begin{equation}
\vec{0}_3\times \vec{0}_{p - 1}\times \mathbb{R}^{d - p - 3}.
\end{equation}
We label the coordinates as in \eqref{coordinatelabels}, and we denote by
\begin{equation}
\Omega_n = \frac{2\pi^{\frac{n + 1}{2}}}{\Gamma(\frac{n + 1}{2})}
\end{equation}
the area of the unit $n$-sphere. The electric and magnetic potentials are
\begin{align}
\phi_E &= -\frac{e}{(d - p - 2)\Omega_{d - p - 1}((\vec{x} - \vec{R})^2 + \vec{z}^2)^{(d - p - 2)/2}}, \nonumber \\
\phi_B &= -\frac{g}{p\Omega_{p + 1}(\vec{x}^2 + \vec{y}^2)^{p/2}}.
\end{align}
The corresponding electric and magnetic fields are
\begin{equation}
\vec{E} = \frac{e\hat{x}_e}{\Omega_{d - p - 1}|\vec{x}_e|^{d - p - 1}}, \qquad \vec{B} = \frac{g\hat{x}_g}{\Omega_{p + 1}|\vec{x}_g|^{p + 1}},
\end{equation}
where we have set
\begin{align}
\vec{x}_e &\equiv (\vec{x} - \vec{R})\times \vec{0}_{p-1}\times \vec{z}, \\
\vec{x}_g &\equiv \vec{x}\times \vec{y}\times \vec{0}_{d - p - 3}.
\end{align}
The constants are important for obtaining the precise normalization factors in the Dirac quantization condition.

Accounting for the dimensions of the branes, the electric potential is a spatial $(p - 1)$-form
\begin{equation}
\Phi_E = \phi_E\, dy_1\wedge \cdots\wedge dy_{p-1},
\end{equation}
and the magnetic potential is a spatial $(d - p - 3)$-form
\begin{equation}
\Phi_B = \phi_B\, dz_1\wedge \cdots\wedge dz_{d - p - 3}.
\end{equation}
We then have
\begin{align}
E &= d\Phi_E = E^i\, (dx_e)_i\wedge dy_1\wedge \cdots\wedge dy_{p-1}, \\
B &= d\Phi_B = B^i\, (dx_g)_i\wedge dz_1\wedge \cdots\wedge dz_{d - p - 3}.
\end{align}
Using the identity
\begin{equation}
\nabla\cdot\left(\frac{\hat{r}}{r^n}\right) = \Omega_n\delta^{n + 1}(\vec{r}), \qquad \vec{r}\in \mathbb{R}^{n + 1},
\end{equation}
we see that
\begin{align}
d\ast'\! E &= (-1)^{p - 1}e\delta^{d-p}(\vec{x}_e)\ast'\!(dy_1\wedge \cdots\wedge dy_{p-1}), \\
d\ast'\! B &= (-1)^{d - p - 3}g\delta^{p+2}(\vec{x}_g)\ast'\!(dz_1\wedge \cdots\wedge dz_{d - p - 3}).
\end{align}
Hence $d\ast' E$ and $d\ast' B$ are proportional to the Poincar\'e duals of the electric and magnetic branes, respectively.

In the angular momentum formula \eqref{formula}, we substitute
\begin{align}
E &= \partial^i\phi_E\, dx_i\wedge dy_1\wedge \cdots\wedge dy_{p-1} + \cdots, \\
B &= \partial^i\phi_B\, dx_i\wedge dz_1\wedge \cdots\wedge dz_{d - p - 3} + \cdots,
\end{align}
where the omitted terms in $E$ involve $dz_i$ and the omitted terms in $B$ involve $dy_i$. For simplicity, we take the separation $\vec{R} = (0, 0, R_3)$ to point in the $x_3$-direction. Then we have
\begin{align}
J_{12} &= (-1)^p\int d^{d-1}x\, ((x_1\partial_1 + x_2\partial_2)\phi_E\partial_3\phi_B \nonumber \\
&\hspace{3 cm} - \partial_3\phi_E(x_1\partial_1 + x_2\partial_2)\phi_B) \label{firstline} \\
&= \frac{(-1)^p eg}{\Omega_{d - p - 1}\Omega_{p + 1}}\int d^{d-1}x \nonumber \\
&\hspace{1 cm} \frac{R_3(x_1^2 + x_2^2)}{((\vec{x} - \vec{R})^2 + \vec{z}^2)^{(d - p)/2}(\vec{x}^2 + \vec{y}^2)^{(p + 2)/2}}.
\end{align}
By dimensional analysis, this integral is independent of $R_3$. Now we evaluate the integrals over $\vec{y}$ and $\vec{z}$ recursively using
\begin{equation}
\int_{-\infty}^\infty \frac{dx}{(a + x^2)^{p/2}} = \frac{1}{a^{(p - 1)/2}}\frac{\Omega_{p - 1}}{\Omega_{p - 2}},
\label{recursive}
\end{equation}
which holds for $p > 1$ and $a > 0$ (the $a$-dependence follows from rescaling $x$). Note that it is valid to apply \eqref{recursive} inside the integral because $a > 0$ almost everywhere. This gives
\begin{equation}
J_{12} = \frac{(-1)^p eg}{16\pi^2}\int d^3 x\, \frac{R_3(x_1^2 + x_2^2)}{((\vec{x} - \vec{R})^2)^{3/2}(\vec{x}^2)^{3/2}},
\end{equation}
which can be evaluated just as in the familiar case of $d = 4$, $p = 1$:
\begin{align}
J_{12} &= -\frac{(-1)^p eg}{16\pi^2}\int d^3 x\, \frac{\vec{x} - \vec{R}}{((\vec{x} - \vec{R})^2)^{3/2}}\cdot \nabla\left(\frac{x_3}{|\vec{x}|}\right) \nonumber \\
&= \frac{(-1)^p eg}{4\pi}\hat{R}_3. \label{eq:generalres}
\end{align}
The result \eqref{eq:generalres} encompasses our earlier result \eqref{4dresult} for $d = 4$, $p = 1$, as well as that derived in \cite{Deser:1997se} for the dyonic case $d = 2(p + 1)$.\footnote{The proof in \cite{Deser:1997se} proceeds by integrating the second term of \eqref{firstline} by parts to transfer the $x_3$-derivative to $\phi_B$ and the $x_{1, 2}$-derivatives to $\phi_E$, leading to partial cancellation with the first term:
\begin{equation}
J_{12} = -2(-1)^p\int d^{d-1}x\, \phi_E\partial_3\phi_B.
\end{equation}
After recursively applying \eqref{recursive}, we obtain
\begin{equation}
J_{12} = -\frac{(-1)^p eg}{8\pi^2}\int d^3 x\, \frac{1}{((\vec{x} - \vec{R})^2)^{1/2}}\partial_3\frac{1}{(\vec{x}^2)^{1/2}},
\end{equation}
which is equivalent to our expression but less amenable to exact evaluation.} It matches \eqref{defpairwise} exactly, where the first term of \eqref{defpairwise} in the case that $d = 2(p + 1)$ follows from the exchange (anti)symmetry of the $p$-forms $E$ and $B$ in the formula \eqref{formula}, so that $q_{ij} = (-1)^p q_{ji}$. As noted in \cite{Deser:1997se}, this argument automatically produces the sign $(-1)^p$ in the DSZ quantization condition \eqref{DSZ}.

To summarize, we have identified the pairwise helicities $q_{ij}$ defined in \eqref{defpairwise} with the angular momentum in the electromagnetic field sourced by dual branes in any $d\geq 4$. Hence the little group transformation of multi-brane states involves an extra phase of $e^{i\sum_{ij} q_{ij}\phi_{ij}}$, just as for electric-magnetic multiparticle states in 4D \cite{Csaki:2020yei}.

\section{\texorpdfstring{$p$}{p}-Form Electrodynamics} \label{app:pform}

We summarize here some additional useful facts about $p$-form electrodynamics. Note that locality constrains $p$-\linebreak[1]form gauge theories with $p > 1$ to be abelian \cite{Teitelboim:1985ya}.\footnote{In modern language, these theories arise from gauging $(p - 1)$-form global symmetries \cite{Gaiotto:2014kfa}, which are abelian for $p - 1 > 0$ because charge operators of codimension $> 1$ have no canonical ordering.}

\subsection{Classical Action}

Given a brane with $p$-dimensional worldvolume $V$ and charge density $\mu$, we define its $p$-form brane current $J$ by
\begin{equation}
\ast J = \mu\widehat{V},
\end{equation}
with $\widehat{V}$ being the Poincar\'e dual of $V$ in spacetime.  In components, we have
\begin{equation}
J^{\mu_1\cdots \mu_p}(x) = \mu\int_V \delta^d(x - X(\sigma))\, dX(\sigma)^{\mu_1}\wedge \cdots\wedge dX(\sigma)^{\mu_p},
\end{equation}
where $x^\mu$ are spacetime coordinates, $X^\mu$ are embedding coordinates, and $\sigma^1, \ldots, \sigma^p$ are coordinates internal to $V$.

The action of $p$-form electrodynamics is obtained by coupling the Maxwell action of a $p$-form gauge field $A$ to electric and magnetic sources.  To write an action that yields the desired field equations
\begin{equation}
dF = \ast J_m, \qquad d\ast F = \ast J_e,
\end{equation}
we use a single singular gauge potential defined on all of spacetime rather than multiple regular potentials defined on patches of spacetime \cite{Teitelboim:1985yc, Henneaux:1986ht}.  For simplicity, we consider a single electric $(p - 1)$-brane and a single magnetic $(d - p - 3)$-brane (the generalization to multiple branes is obvious).  We denote their brane currents by
\begin{equation}
\ast J_e = e\widehat{V_e}, \qquad \ast J_m = g\widehat{V_m},
\end{equation}
respectively.  To allow for magnetic sources without violating the Bianchi identity $d^2 A = 0$, we further introduce a $(d - p - 2)$-dimensional Dirac brane ending on the magnetic brane and a $(d - p - 1)$-form current $G$ localized to the Dirac brane worldvolume satisfying
\begin{equation}
d\ast G = \ast J_m.
\end{equation}
We then define the field strength
\begin{equation}
F\equiv dA + \ast G,
\end{equation}
in terms of which the classical action is
\begin{equation}
S[A, G] = -\frac{1}{2}\int F\wedge\ast F - (-1)^p\int A\wedge\ast J_e.
\end{equation}
In natural units, the form fields $A$ and $F$ and the charge densities $e$ and $g$ have mass dimensions
\begin{equation}
[A] = [F] = -[e] = [g] = \frac{d}{2} - (p + 1).
\end{equation}
This action treats electric and magnetic sources highly asymmetrically.  While the coupling of $A$ to the electric brane is manifest, the coupling of $A$ to the magnetic brane (via the Dirac brane) is hidden in the $F\wedge\ast F$ term.  To ensure that the Dirac brane introduces no independent dynamics (i.e., that the equation of motion for $G$ follows from that of $A$), we require that the Dirac brane does not intersect the electric brane (``Dirac veto'').

We assume that the brane sources are non-dynamical.  To make the branes dynamical, we would add a kinetic term for each brane of the form
\begin{equation}
-T\int_V \ast 1 = -T\int_V d^{\dim V}\sigma\sqrt{|h|},
\end{equation}
where $\ast 1$ is the appropriate volume form on the worldvolume $V$ (with coordinates $\sigma$) and the tension $T$ has mass dimension $[T] = \dim V$.  The signature of the induced metric $h_{ab} = \partial_{\sigma^a}X_\mu\partial_{\sigma^b}X^\mu$ depends on the position of the brane.  The action would then additionally be a functional of the brane embedding coordinates $X(\sigma)$.

Both fields $A$ and $G$ are singular along the Dirac brane, while $F$ is nonsingular except at the magnetic brane.  The Dirac brane current $G$ cancels the magnetic flux that would otherwise be carried away from the magnetic brane by the singular part of the field due to $A$.  The Dirac brane is unphysical because its configuration is gauge-dependent.  To see this, note that if spacetime is topologically trivial, then the existence of a Dirac brane worldvolume $V_D$ satisfying $\partial V_D = V_m$ follows from the fact that the magnetic brane worldvolume $V_m$ is closed ($\partial V_m = 0$).  This relation determines only the homology class of $V_D$, so the Dirac brane is ambiguous up to
\begin{equation}
V_D\to V_D + \partial V.
\end{equation}
Via \eqref{morphism}, we may take $\ast G = (-1)^p g\widehat{V_D}$, which then satisfies $d\ast G = \ast J_m$ because $\partial V_D = V_m$.  This ambiguity translates into a gauge freedom
\begin{equation}
\ast G\to \ast G + d\ast\Lambda, \qquad A\to A + d\lambda - \ast\Lambda,
\end{equation}
where $\ast\Lambda = -g\widehat{V}$.  The gauge field $A$ must undergo a compensatory gauge transformation to leave $F$ invariant.  $\ast\Lambda$ represents a \emph{singular} gauge transformation of $A$ (on top of ordinary gauge transformations $A\to A + d\lambda$) because $\ast\Lambda$ is only locally exact.

The field equations are then the modified Bianchi identity for $F$ and the equation of motion for $A$, with variations taken from the left.  Both $J_e$ and $J_m$ are conserved, as required by gauge invariance.  Conservation of brane current ($d\ast J = 0$) is Poincar\'e dual to the statement that the worldvolume has no boundary ($\partial V = 0$).

\subsection{Classical Duality Group}

When $d = 2(p + 1)$, dyonic branes exist, the charge density is dimensionless, and the $p$-form Maxwell action is conformal: $T_\mu{}^\mu = (d - 2(p + 1))\mathcal{L} = 0$.  In this case, the field equations have the same form degree,
\begin{equation}
d\left(\begin{array}{c} F \\ \ast F \end{array}\right) = \ast\left(\begin{array}{c} J_m \\ J_e \end{array}\right),
\end{equation}
and are therefore preserved by linear transformations
\begin{equation}
\left(\begin{array}{c} F \\ \ast F \end{array}\right)\to R\left(\begin{array}{c} F \\ \ast F \end{array}\right), \quad \left(\begin{array}{c} J_m \\ J_e \end{array}\right)\to R\left(\begin{array}{c} J_m \\ J_e \end{array}\right)
\end{equation}
for $R\in GL(2, \mathbb{R})$ \cite{Deser:1997mz}.  Further requiring such a transformation to respect the Hodge duality condition $\ast^2 F = (-1)^p F$ and to preserve the stress tensor restricts $R$ to lie in $SO(2)$ for $p$ odd and $\mathbb{Z}_2\times \mathbb{Z}_2$ for $p$ even, where the $\mathbb{Z}_2$ factors act as $F\leftrightarrow -F$ and $F\leftrightarrow \ast F$.  This is the classical electric-magnetic duality group of $p$-form electrodynamics.  It can be promoted to an off-shell symmetry of the action using a two-potential formulation \cite{Deser:1997mz} in which
\begin{equation}
\left(\begin{array}{c} F \\ \ast F \end{array}\right) = d\left(\begin{array}{c} A \\ B \end{array}\right) + \ast\left(\begin{array}{c} G \\ H \end{array}\right),
\end{equation}
where $A, B$ are the electric and magnetic potentials and $G, H$ are Dirac brane currents for the magnetic and electric branes satisfying $d\ast G = \ast J_m$ and $d\ast H = \ast J_e$.  We can then imagine the usual minimal coupling to the brane worldvolumes:
\begin{equation}
-(-1)^p e\int_{V_e} A - (-1)^{d - p - 2}g\int_{V_m} B.
\end{equation}
Such an action sacrifices manifest Lorentz invariance. The construction of a Lagrangian for both electric and magnetic charges that simultaneously manifests locality and Lorentz invariance is a long-standing problem \cite{Dirac:1948um, Zwanziger:1970hk}.

The electric-magnetic duality group can be extended to $SL(2, \mathbb{R})$ in the presence of a theta term for $d = 2(p + 1)$ and $p$ odd.\footnote{``Diagonal'' theta terms are only allowed when $4|d$ since $F\wedge F = (-1)^{p + 1}F\wedge F$ by antisymmetry of the wedge product.  However, mixed theta terms are allowed for $p$ even \cite{Deser:1998vc}.  We do not discuss the inclusion of Chern-Simons terms in odd $d$ \cite{Henneaux:1986ht}.}  To see this, we rescale $A\to e^{-1}A$ so that the charge densities become $e\to 1$ and $g\to g/e$, resulting in a non-canonically normalized kinetic term:
\begin{equation}
S = -\frac{1}{2e^2}\int F\wedge\ast F + \frac{\theta}{4\pi}\int F\wedge F - (-1)^p\int_V A,
\end{equation}
where all couplings are dimensionless.  Assuming that the Dirac branes do not intersect ($\ast G\wedge\ast G = 0$) and dropping boundary terms, we can write
\begin{equation}
S = -\frac{1}{2e^2}\int F\wedge\ast F - (-1)^p\int A\wedge\left(\ast J_e + \frac{\theta}{2\pi}\ast J_m\right).
\end{equation}
The field equations are modified to
\begin{equation}
d\left(\begin{array}{c} F \\ \frac{1}{e^2}\ast F \end{array}\right) = \ast\left(\begin{array}{c} J_m \\ J_e + \frac{\theta}{2\pi}J_m \end{array}\right),
\end{equation}
thus manifesting the Witten effect (the theta angle shifts the electric charge of a dyonic brane by $\frac{\theta g}{2\pi}$) \cite{Deser:1997se, Deser:1998vc}.  In terms of the complex coupling and the complex field strength
\begin{equation}
\tau = \tau_1 + i\tau_2 = \frac{\theta}{2\pi} + \frac{i}{e^2}, \qquad F^c = F + i\ast F,
\end{equation}
and defining the two-component objects
\begin{equation}
\psi\equiv \left(\begin{array}{c} 1 \\ -\tau \end{array}\right), \quad \mathcal{F}\equiv \operatorname{Re}(\psi F^c), \quad \mathcal{J}\equiv \left(\begin{array}{c} J_m \\ J_e \end{array}\right),
\end{equation}
the field equations become
\begin{equation}
d\mathcal{F} = \ast\mathcal{J}.
\end{equation}
They are invariant under
\begin{equation}
\psi\to (R^{-1})^T\psi, \qquad \mathcal{J}\to (R^{-1})^T\mathcal{J}
\end{equation}
for $R\in GL(2, \mathbb{R})$.  The requirement of preserving the self-duality condition
\begin{equation}
\mathcal{F} = \gamma\omega\ast \mathcal{F}, \quad \gamma\equiv \frac{1}{\tau_2}\left(\begin{array}{cc} 1 & -\tau_1 \\ -\tau_1 & |\tau|^2 \end{array}\right), \quad \omega\equiv \left(\begin{array}{cc} 0 & -1 \\ 1 & 0 \end{array}\right)
\end{equation}
in $d = 2(p + 1)$ with $p$ odd restricts $R\in SL(2, \mathbb{R})$.  Indeed, writing
\begin{equation}
\gamma = \frac{\psi\psi^\dag + \text{c.c.}}{\sqrt{\det(\psi\psi^\dag + \text{c.c.})}}
\end{equation}
(following \cite{Gibbons:1995cv}), we see that the $SL(2, \mathbb{R})$ transformation
\begin{equation}
R = \left(\begin{array}{cc} a & b \\ c & d \end{array}\right), \qquad ad - bc = 1
\end{equation}
takes $\gamma\to (R^{-1})^T\gamma R^{-1}$ while $R^{-1}\omega(R^{-1})^T = \omega$.  Correspondingly, the complex coupling transforms as
\begin{equation}
\tau\to \frac{a\tau + b}{c\tau + d}.
\end{equation}

\subsection{Dirac Quantization Condition}

The Dirac quantization condition (DQC) follows by demanding invariance under singular gauge transformations \cite{Bekaert:2002cz}.  The change in the action under a singular gauge transformation comes from the coupling of $A$ to $J_e$:
\begin{equation}
\Delta S\propto \int \ast\Lambda\wedge\ast J_e\propto egL(V_e, \partial V),
\end{equation}
where the proportionality factors are signs.  This immediately gives the DQC, as long as the variation in the Dirac brane worldvolume can link with the electric brane worldvolume.

Now consider dimensions $d = 2(p + 1)$.  Consider two dyonic branes of charge densities $(e_1, g_1)$ and $(e_2, g_2)$ as well as worldvolumes and corresponding Dirac branes $V_1 = \partial D_1$ and $V_2 = \partial D_2$.  A variation in the configurations of the Dirac branes,
\begin{equation}
D_1\to D_1'\equiv D_1 + \partial E_1, \quad D_2\to D_2'\equiv D_2 + \partial E_2,
\end{equation}
is implemented by a singular gauge transformation that takes
\begin{equation}
A\to A - \ast\Lambda, \qquad \ast\Lambda = -g_1\widehat{E_1} - g_2\widehat{E_2}.
\end{equation}
Again, the change in the action comes from the coupling of $A$ to $J_e$ (the theta term, if present, does not affect the argument because it is manifestly gauge-invariant).  Using $\ast J_e = e_1\smash{\widehat{V_1}} + e_2\smash{\widehat{V_2}}$, we can write this variation as an intersection number:
\begin{align}
\Delta S &= \textstyle (-1)^p\int \ast\Lambda\wedge\ast J_e \\
&= -(e_1 g_2 I(V_1, E_2) + e_2 g_1 I(V_2, E_1)) \\
&= -(e_1 g_2 I(\partial D_1, E_2) + e_2 g_1 I(\partial D_2, E_1)) \\
&= -(e_1 g_2 I(\partial E_2, D_1) + e_2 g_1 I(\partial E_1, D_2)),
\end{align}
where we use that $\dim V_i = \dim D_i - 1 = \dim E_i - 2 = p$.  For arbitrary variations of the Dirac branes, we clearly require $e_1 g_2\in 2\pi\mathbb{Z}$ and $e_2 g_1\in 2\pi\mathbb{Z}$ separately (the DQC), but for variations in which the Dirac branes in their initial and final configurations do not intersect, we have
\begin{equation}
I(D_1, D_2) = I(D_1', D_2') = 0.
\end{equation}
The intersection number of two boundaries is zero since $\partial^2 = 0$, so $I(\partial E_1, \partial E_2) = I(D_1' - D_1, D_2' - D_2) = 0$, from which we see that
\begin{equation}
I(D_1, D_2') + I(D_1', D_2) = 0.
\end{equation}
Hence we deduce that
\begin{equation}
\Delta S = (-1)^p(e_1 g_2 + (-1)^p e_2 g_1)I(D_1, D_2'),
\end{equation}
and we get $e_1 g_2 + (-1)^p e_2 g_1\in 2\pi\mathbb{Z}$.  The consistency of this DSZ quantization condition with dimensional reduction is discussed in \cite{Bremer:1997qb, Deser:1998vc}.

\subsection{Quantum Duality Group}

The DSZ pairing appearing in \eqref{DSZ} is invariant under the action of the classical electric-magnetic duality group, namely $SL(2, \mathbb{R})$ for $p$ odd and the $\mathbb{Z}_2\times \mathbb{Z}_2$ generated by $\{\left(\begin{smallmatrix} 0 & 1 \\ 1 & 0 \end{smallmatrix}\right), \left(\begin{smallmatrix} -1 & 0 \\ 0 & -1 \end{smallmatrix}\right)\}$ for $p$ even.  In general, however, only a discrete subgroup of the classical duality group preserves the quantum-mechanical charge lattice.

In $d = 2(p + 1)$ with $p$ odd, \eqref{DSZ} takes the form
\begin{equation}
e_1 g_2 - e_2 g_1\in 2\pi\mathbb{Z}.
\end{equation}
The relative minus sign ensures that the quantization condition is invariant under the Witten effect.  In a convenient normalization, the maximal charge lattice allowed by \eqref{DSZ} is
\begin{equation}
(e, g) = (m + \theta n, 2\pi n)
\end{equation}
for $m, n\in \mathbb{Z}$.  Letting $q = e + ig\in \mathbb{C}$, this lattice defines a torus $q\sim q + 1$ and $q\sim q + 2\pi\tau$ with $\tau = \frac{\theta}{2\pi} + i$.  The subgroup of the classical duality group $SL(2, \mathbb{R})$ that preserves the charge lattice is $SL(2, \mathbb{Z})$.  If $\theta$ is constrained to vanish, then the subgroup of the classical duality rotation group $SO(2)$ that preserves the resulting rectangular charge lattice is $\mathbb{Z}_4$.

In $d = 2(p + 1)$ with $p$ even, \eqref{DSZ} takes the form
\begin{equation}
e_1 g_2 + e_2 g_1\in 2\pi\mathbb{Z},
\end{equation}
and theta terms are not allowed.  The corresponding charge lattice $(e, g) = (m, 2\pi n)$ for $m, n\in \mathbb{Z}$ is preserved by the full classical duality group $\mathbb{Z}_2\times \mathbb{Z}_2$.

The $SL(2, \mathbb{Z})$ duality is an ambiguity in the presentation of the theory, i.e., in the choice of which sources to call electric or magnetic.  It can be thought of as a change of variables in the path integral.  Of the generators of $SL(2, \mathbb{Z})$, the $T$-transformation $T : \tau\mapsto \tau + 1$ simply takes $\theta\to \theta + 2\pi$.  The $S$-transformation $S : \tau\mapsto -1/\tau$ also has a simple path integral interpretation.  For $d = 2(p + 1)$ with $p$ odd, consider the rescaled action with no sources:
\begin{equation}
S[A] = \int \left(-\frac{1}{2e^2}F\wedge\ast F + \frac{\theta}{4\pi}F\wedge F\right), \quad F\equiv dA.
\end{equation}
Assuming that spacetime is topologically trivial, we can instead view the action as a functional of the fundamental fields $B$ and $F$, where $B$ is a $p$-form Lagrange multiplier that enforces the Bianchi identity, thus implying that $F$ is the field strength of a $p$-form potential:
\begin{equation}
S[B, F] = \int \left(-\frac{1}{2e^2}F\wedge\ast F + \frac{\theta}{4\pi}F\wedge F - B\wedge dF\right).
\end{equation}
Integrating out $F$ then gives
\begin{equation}
S[B] = \int \left(-\frac{1}{2e'^2}G\wedge\ast G + \frac{\theta'}{4\pi}G\wedge G\right), \quad G\equiv dB,
\end{equation}
where the new complex coupling is
\begin{equation}
\tau' = \frac{\theta'}{2\pi} + \frac{i}{e'^2} = -\frac{1}{\tau}.
\end{equation}
Under the duality $F\leftrightarrow G$, the equation of motion for $A$ goes to the Bianchi identity for $B$ and vice versa.  A similar argument establishes the quantum equivalence of abelian $p$-form and $(d - p - 2)$-form gauge fields.

\end{document}